\documentclass[12pt]{article} \hoffset=-15mm \voffset=-10mm
\usepackage{graphicx,amsmath,amssymb,epsf} 
\usepackage{latexsym,bm,slashed} 
\usepackage{xcolor} \definecolor{dark}{rgb}{0.10,0.2,0.3}
\definecolor{magenta}{rgb}{0.7,0.1,0.3}
\definecolor{purpure}{rgb}{0.5,0.15,0.3}
\usepackage[font=small,format=plain,labelfont=bf,up,textfont=it,up]{caption}
\usepackage{hyperref, cite} \hypersetup{colorlinks, citecolor=blue,
  filecolor=blue, linkcolor=magenta,
  urlcolor=purpure,hyperfootnotes=true,pdftex}

\def\be{\begin{equation}}
\def\ee{\end{equation}}
\def\bea{\begin{eqnarray}}
\def\eea{\end{eqnarray}}

\def\ubar{\bar{u}}

\newcommand{\slv}{\raise.15ex\hbox{$/$}\kern-.53em\hbox{$v$}}
\newcommand{\sln}{\raise.15ex\hbox{$/$}\kern-.53em\hbox{$n$}}
\newcommand{\slnbar}{\raise.15ex\hbox{$/$}\kern-.53em\hbox{$\bar{n}$}}
\newcommand{\slF}{\raise.15ex\hbox{$/$}\kern-.53em\hbox{$F$}}
\newcommand{\sll}{\raise.15ex\hbox{$/$}\kern-.40em\hbox{$l$}}
\newcommand{\sllbar}{\raise.15ex\hbox{$/$}\kern-.40em\hbox{$\bar{l}$}}
\newcommand{\slh}{\raise.15ex\hbox{$/$}\kern-.40em\hbox{$h$}}
\newcommand{\slP}{\raise.15ex\hbox{$/$}\kern-.53em\hbox{$P$}}
\newcommand{\slp}{\raise.15ex\hbox{$/$}\kern-.53em\hbox{$p$}}
\newcommand{\slq}{\raise.15ex\hbox{$/$}\kern-.53em\hbox{$q$}}
\newcommand{\slqbar}{\raise.15ex\hbox{$/$}\kern-.53em\hbox{$\bar{q}$}}
\newcommand{\slR}{\raise.15ex\hbox{$/$}\kern-.53em\hbox{$R$}}
\newcommand{\slz}{\raise.15ex\hbox{$/$}\kern-.53em\hbox{$Z$}}
\newcommand{\slzbar}{\raise.15ex\hbox{$/$}\kern-.53em\hbox{$\bar{Z}$}}
\newcommand{\slQ}{\raise.15ex\hbox{$/$}\kern-.53em\hbox{$Q$}}
\newcommand{\slK}{\raise.15ex\hbox{$/$}\kern-.53em\hbox{$K$}}
\newcommand{\slk}{\raise.15ex\hbox{$/$}\kern-.53em\hbox{$k$}}
\newcommand{\slkbar}{\raise.15ex\hbox{$/$}\kern-.53em\hbox{$\bar{k}$}}
\newcommand{\slkone}{\raise.15ex\hbox{$/$}\kern-.53em\hbox{$k_1$}}
\newcommand{\slpone}{\raise.15ex\hbox{$/$}\kern-.53em\hbox{$p_1$}}
\newcommand{\slpbarone}{\raise.15ex\hbox{$/$}\kern-.53em\hbox{$\bar{p}_1$}}
\newcommand{\slptwo}{\raise.15ex\hbox{$/$}\kern-.53em\hbox{$p_2$}}
\newcommand{\slpbartwo}{\raise.15ex\hbox{$/$}\kern-.53em\hbox{$\bar{p}_2$}}
\newcommand{\slqone}{\raise.15ex\hbox{$/$}\kern-.53em\hbox{$q_1$}}
\newcommand{\slD}{\raise.15ex\hbox{$/$}\kern-.53em\hbox{$\!D$}}
\newcommand{\slC}{\raise.15ex\hbox{$/$}\kern-.53em\hbox{$C$}}
\newcommand{\slA}{\raise.15ex\hbox{$/$}\kern-.73em\hbox{$A$}}
\newcommand{\slSigma}{\raise.15ex\hbox{$/$}\kern-.53em\hbox{$\Sigma$}}
\newcommand{\slpartial}{\raise.15ex\hbox{$/$}\kern-.53em\hbox{$\partial$}}
\newcommand{\slcalP}{\raise.15ex\hbox{$/$}\kern-.63em\hbox{$\cal P$}}
\newcommand{\sleps}{\raise.15ex\hbox{$/$}\kern-.53em\hbox{$\epsilon$}}
\newcommand{\slepsbar}{\raise.15ex\hbox{$/$}\kern-.53em\hbox{$\overline{\epsilon}$}}
\newcommand{\slepsstar}{\raise.15ex\hbox{$/$}\kern-.53em\hbox{$\epsilon$}^\star}
\newcommand{\slS}{\raise.15ex\hbox{$/$}\kern-.73em\hbox{$S$}}

\newcommand{\nn}{\nonumber\\}

 \title{\bf
  \Large \bf \Large Rapidity loss, spin and angular asymmetries in scattering of a quark from color field of a proton (nucleus)} \author{ Jamal~Jalilian-Marian$^{1,2}$
  \bigskip \\
  {\normalsize $^1$Department of Natural Sciences, Baruch College,
    CUNY,}
  \\
  {\normalsize 17 Lexington Avenue, New York, NY 10010, USA}\\
  {\normalsize $^2$CUNY Graduate Center, 365 Fifth Avenue, New York, NY 10016, USA}\\
   }

\begin{document}

\maketitle
\begin{abstract}
We calculate the helicity amplitudes for scattering of a quark from both large and small $x$ gluons of a target proton or nucleus using spinor helicity formalism. We show that scattering from large $x$ gluons of the target results in non-zero spin asymmetry at intermediate $p_t$ as well as rapidity loss of the projectile quark. We comment on how this can also generate angular asymmetries in particle production in high energy collisions.
\end{abstract}

\section{Introduction}
The Color Glass Condensate (CGC) is an effective theory of small $x$ gluons in a proton or nucleus, valid in the limit $x \rightarrow 0$ where QCD cross sections are dominated by small $x$ kinematics and gluon saturation is expected to be the dominant dynamics~\cite{cgc-reviews}. Whereas there are strong and tantalizing hints about presence and contribution of gluon saturation dynamics to forward rapidity particle production and di-jet angular asymmetry in proton-nucleus collisions at RHIC and the LHC, an unambiguous and definitive interpretation of the data in terms of gluon saturation is still lacking~\cite{eic}. To help clarify the contribution of gluon saturation to these processes, higher order (in $\alpha_s$) corrections~\cite{nlo} to various particle production cross sections have been computed which improve the precision of CGC calculations in the small $x$ kinematics, commonly taken to be $x \le 0.01$. 
 
While higher order corrections are invaluable for precision studies of CGC, they are still limited to the small $x$ kinematics where $x \le 0.01$. Recalling the kinematic relation between transverse momentum and rapidity of a produced particle in high energy collisions and the Bjorken $x$ of the gluons of the target probed in the scattering given by $x \sim {p_t \over \sqrt{s}}\, e^{- y}$ we see that production of high $p_t$ particles is necessarily dominated by large $x$ kinematics. Since the center of mass energy $\sqrt{s}$ of a high energy collision, such as that at RHIC or the LHC, is fixed it is perhaps more useful to think of small vs large $x$ limits of the observables rather than the $\sqrt{s} \rightarrow \infty$ limit as is commonly done in CGC formalism. This is also important because  the large $x$ kinematics will be a significant part of the phase space of proton/nucleus wave function probed in all proposed Electron Ion Colliders~\cite{eic}. For the first time it will be possible to  experimentally investigate in detail the transition from large to small $x$ dynamics in large nuclei in a transverse momentum region where genuinely non-perturbative QCD effects may not be significant.   

Even more significant is perhaps the common estimates of the $x$ kinematics contributing to a production cross section in CGC formalism. Unlike DIS structure functions which can be measured at various $Q^2$ at a fixed value of $x$, particle production in proton (nucleus)-proton (nucleus) collisions in the collinear factorization formalism involves a convolution in $x$ and is sensitive to a range of parton $x$ in the target (and projectile). In the small $x$ approximation employed in CGC calculations one assumes the target gluon distribution $xG (x, Q^2)$ is rising so fast with $1/x$ that one can approximate it as being given by the minimum value of x ($\equiv x_{min}$) in the convolution, symbolically written as            
\be
\int^1_{x_{min}} d x \, x G (x, Q^2) \cdots \simeq \, x_{min} G (x_{min}, Q^2) 
\ee
where $\cdots$ stands for the rest of the collinearly factorized cross section. While this kind of an approximation may be fine for making parametric estimates or even for semi-quantitative analysis of the data it can not be expected to be precise as the above approximation disregards contribution of the larger $x$ ($x \ge 0.01$) kinematics~\footnote{For a comparison of the target $x$ kinematics contributing to a given process, compare Fig. ($10$) in \cite{dhj} with Fig. ($1$) in \cite{gsv}.}. Furthermore making the small $x$ approximation allows one to use eikonal methods~\cite{eik-reviews} which treat the projectile parton as moving on a straight line and not deflected after multiply scattering from the target. Clearly this can not be a good approximation when the scattered parton is at high $p_t$, i.e. deflected by a large angle.        

In \cite{jjm-largex} we proposed a new approach which aims to include contribution of both small and large $x$ partons of the target to a scattering cross section (see also \cite{martin}). The small $x$ gluons of the target are treated via CGC methods while large $x$ gluons of the target are treated as the standard partons of QCD-improved parton model. To do so we calculated the amplitude for multiple scatterings of a quark from a color field $\mathcal{A}^\mu = S^\mu + A^\mu$ where $A^\mu = (\mathcal{A}^\mu - S^\mu)$ and
$S^\mu = \delta^{\mu \, -}$ in light cone gauge. We included multiple scatterings from the soft field $S^\mu$, radiated by small $x$ gluons, to all orders but kept only the single scattering contribution from $A^\mu$ field radiated by a large $x$ gluon. This allows exchange of potentially large longitudinal and transverse momentum between the projectile and the target, unlike the CGC formalism where scattering involves exchange of small transverse momenta only and no longitudinal momentum is exchanged. 

Here we use spinor helicity formalism~\cite{dixon} to calculate the helicity amplitudes for scattering of a quark on a proton or nucleus target, including both small and large $x$ gluons of the target. We show that the scattering cross section is sensitive to the helicity of the projectile quark and that inclusion of scattering from the large $x$ gluon results in spin asymmetry for quark scattering. Furthermore, there is longitudinal momentum transferred from the projectile to the target which results in projectile rapidity loss, unlike the CGC formalism where the projectile longitudinal momentum 
remains unchanged. We argue that, coupled to a realistic description of the target geometry, this would also lead to angular asymmetries in particle production in high energy proton-proton (nucleus) collisions.   

\section{Helicity amplitudes}

\begin{figure}[h]
  \centering
  \includegraphics[width=0.9\textwidth]{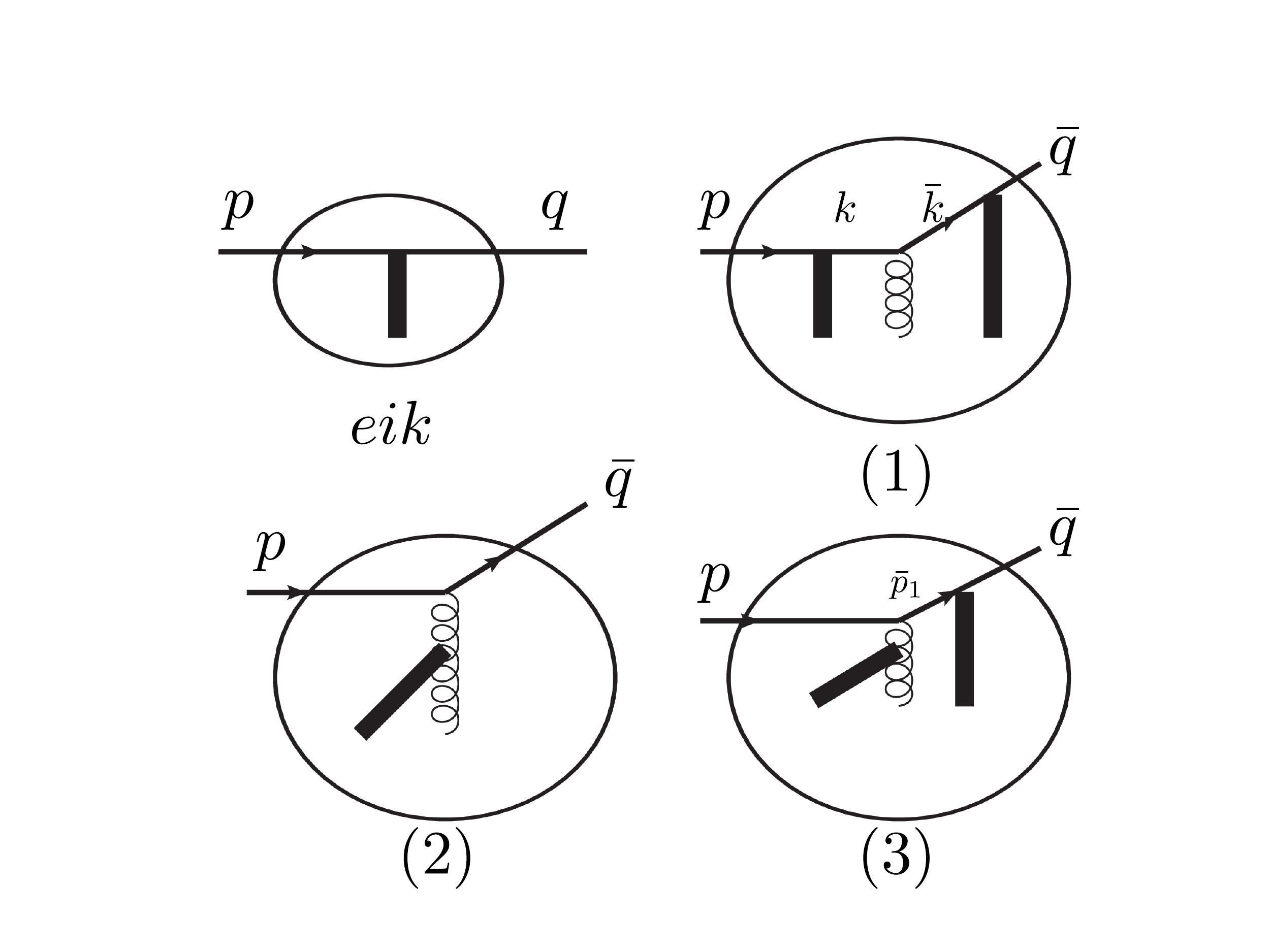}
  \caption{\it Scattering of a quark from small and large $x$ gluons of the target.}
  \label{fig:allfigs}
\end{figure}

The amplitude for scattering of a quark projectile on both small and large $x$ gluons of a target proton or nucleus was computed in \cite{jjm-largex}. Here we use spinor helicity methods to evaluate the amplitude for a given projectile quark helicity. There are in principle $4$ classes of diagrams contributing as shown in Fig. (\ref{fig:allfigs}). 
Multiple soft scatterings of the quark (gluon) from the classical color field representing 
small $x$ gluons of the target is summed into a semi-infinite Wilson line in fundamental 
(adjoint) representation which is shown as solid black rectangles while the wavy line 
represents single scattering from a large $x$ gluon of the target proton or nucleus , which 
itself is shown as a big open ellipse. The eikonal diagram, labeled "eik" above, is the standard diagram used in the hybrid formulation of 
quark scattering from the small $x$ gluons of a target in CGC approach~\cite{hybrid}. The other three diagrams, labeled $1 - 3$, are non-zero only in the large $x$ kinematics and vanish in the small $x$ limit. The amplitudes are (for details and definitions of Wilson lines see \cite{jjm-largex}, note that anti-path-ordering label is dropped for the sake of compactness.),

\bea
i \mathcal{M}_{eik} (p,q) &=& 2 \pi \delta (p^+ - q^+)\,  
\int d^2 x_{t}\, e^{- i (q_t - p_t) \cdot x_{t}} \, 
\left[V (x_t) - 1\right]\, \mathcal{N}_{eik} \nn
i \mathcal{M}_1 (p,q) &=&  \int d^4 x\, d^2 z_t \, d^2 \bar{z}_t \, 
\int {d^2 k_t \over (2 \pi)^2} \, {d^2 \bar{k}_t \over (2 \pi)^2} \, 
e^{i (\bar{k} - k) x} \, 
e^{- i (\bar{q}_t - \bar{k}_t)\cdot \bar{z}_t}\, 
e^{- i (k_t - p_t)\cdot z_t}
 \nn
&&
\overline{V} (x^+, \bar{z}_t) \, (i g\, t^b) \, V (z_t, x^+) \,  \mathcal{N}_1^b\nn 
i \mathcal{M}_2 (p,q) &=& 2\, i \, \int d^4 x\, 
e^{i (\bar{q}^+ - p^+) x^-  - i (\bar{q}_t - p_t) \cdot x_t}\, (i g \, t^a)\, 
\left[\partial_{x^+}\, U^\dagger (x_t, x^+)\right]^{a b}\, \mathcal{N}_2^b \nn 
i \mathcal{M}_3 (p,q) &\!\!=\!\!&  - 2\, i\, \int d^4 x\,  d^2 \bar{x}_t \, d \bar{x}^+\,  
 {d^2 \bar{p}_{1 t} \over (2 \pi)^2} \, 
e^{i (\bar{p}_1^+ - p^+) x^-} \, 
e^{- i (\bar{p}_{1 t} - p_t)\cdot x_t}\, 
e^{- i (\bar{q}_t - \bar{p}_{1 t})\cdot \bar{x}_t}
 \nn
&&
\left[\partial_{\bar{x}^+}\, \overline{V} (\bar{x}^+, \bar{x}_t)\right]\, 
(i g\, t^a)\, 
\left[\partial_{x^+}\, U^\dagger (x_t, x^+)\right]^{a b}\, \mathcal{N}_3^b 
\label{eq:iM}
\eea 
where $p, q$ are the momenta of the incoming and outgoing quark. The Dirac structure 
of the numerators in eq. (\ref{eq:iM}) are labeled as $\mathcal{N}$,  respectively 
defined as 
\bea
\mathcal{N}_{eik} &=& \ubar (q)\, \sln\, u(p) \nn
\mathcal{N}_1^b &=& {1 \over 2 k^+} {1 \over 2 \bar{k}^+}\,  
\ubar (\bar{q})\, \left[ \slnbar \, \slkbar  \,
\slA^b (x) \, 
\slk  \, \sln \right]\,  u(p) \nn
\mathcal{N}_2^b &=& {1 \over (p - \bar{q})^2} 
\ubar (\bar{q})\, 
\left[n \cdot (p - \bar{q}) \, \slA^b (x) -  (p - \bar{q}) \cdot A^b (x) \, \sln\right]\, 
\bigg]\, u(p) \nn
\mathcal{N}_3^b &=& 
{1 \over 2 \bar{n} \cdot \bar{p}_1 (p - \bar{p}_1)^2 }
\ubar (\bar{q}) \!\left[
 \slnbar\, \slpbarone \,
\bigg( n \cdot (p - \bar{p}_1) \, \slA^b (x) -  (p - \bar{p}_1) \cdot A^b (x) \, \sln\bigg)
\right]\! u (p)
\label{eq:N}
\eea
Note that we have not written out the color indices in fundamental representation in order to ensure a compact form for the expressions. The contribution of $\mathcal{N}^{\pm}_{eik}$ in the eikonal diagram is $\mathcal{N}^{\pm}_{eik} \sim p^+ \sim \sqrt{s}$ and is already included in CGC calculations. Furthermore it has zero overlap with the new non-ikonal contributions, therefore we will ignore it from now on and focus on the new terms. The specific helicity amplitudes are then\footnote{The superscript $\pm$ refers to helicity of the incoming quark. Since helicity is conserved the helicity of the outgoing quark is the same as the helicity of the incoming quark and is not shown explicitly.} 
\bea
\mathcal{N}_1^{+ , b} &=&  <\bar{k}^+ |\, \slA^b (x) | k^+ > \nn
\mathcal{N}_1^{- , b} &=&  <\bar{k}^- |\, \slA^b (x) | k^- >  \nn
\mathcal{N}_2^{+ , b} &=& {1 \over (p - \bar{q})^2}
\left[n\cdot (p - \bar{q})  <\bar{q}^+ |\, \slA^b (x) | p^+ > -
<\!n p\!> \, [\bar{q} \, n ]\, (p - \bar{q}) \cdot A^b (x) \right]\nn
\mathcal{N}_2^{- , b} &=&
{1 \over (p - \bar{q})^2}
\left[n\cdot (p - \bar{q})  <\bar{q}^- |\, \slA^b (x) | p^- > - 
\, <\!\bar{q} n \!>\, [n\, p] \,(p - \bar{q}) \cdot A^b (x)\right] \nn
\mathcal{N}_3^{+ , b} &=& 
{[\bar{q} \bar{n}] <\!\bar{n} \bar{p}_1\!> 
\!\big[ n\cdot (p - \bar{p}_1)  <\bar{p}_1^+ |\, \slA^b (x) | p^+ > \! - \!
<\!n p\!> [ \bar{p}_1 n ]  (p - \bar{p}_1) \cdot A^b (x) \big] 
\over
2 \bar{n} \cdot \bar{p}_1 (p - \bar{p}_1)^2 
}\nn
\mathcal{N}_3^{- , b} &=& 
{<\!\bar{q} \bar{n}\!> [\bar{n} \bar{p}_1 ] 
\big[ n\cdot (p - \bar{p}_1)  <\bar{p}_1^- |\, \slA^b (x) | p^- > \!-\! 
 <\!\bar{p}_1 n\!> [n p] (p - \bar{p}_1) \cdot A^b (x)\big] 
\over  
2 \bar{n} \cdot \bar{p}_1 (p - \bar{p}_1)^2 
}
\nn
&&
\label{eq:N-hel}
\eea
We note that 
\be
\mathcal{N}_3^{+ , b} = {[\bar{q} \bar{n}] <\!\bar{n} \bar{p}_1\!> 
\over  2 \bar{n} \cdot \bar{p}_1}\, \mathcal{N}_2^{+ , b} (\bar{q} \rightarrow \bar{p}_1)
= 
\mathcal{N}_2^{+ , b} (\bar{q} \rightarrow \bar{p}_1)
\ee
and similarly for $\mathcal{N}_3^{- , b}$. The helicity amplitudes above (\ref{eq:N-hel}) are related via identities like
\bea
<\bar{k}^+ |\, \slA_b \, | k^+>  &=&  < k^- |\, \slA_b \, | \bar{k}^- > \nn
<\bar{k}^+ |\, \slA_b \, | k^+> &=& \left(< \bar{k}^- |\, \slA_b \, | k^- >\right)^\star
\eea
Using these relations we note that 
\be
\mathcal{N}_{1,2,3}^{- , b} = \big[\mathcal{N}_{1,2,3}^{+ , b}\big]^\star
\label{eq:hel-plus-minus}
\ee

\subsection{Evaluating the helicity amplitudes}

To evaluate the helicity amplitudes above we will need the following relations,
\be
<k_1^\pm |\, \gamma^+ \, | k_2^\pm > \, = \, \sqrt{2 k_1^+\, 2 k_2^+} 
\ee
and
\be
<k_1^\pm |\, \gamma_i \, | k_2^\pm > \,  = \, 
\sqrt{k_1^+\, k_2^+} \,
\bigg\{
{k_{1 i} \mp i \epsilon_{i j} \, k_1^j \over k_1^+ }
+ {  k_{2 i} \pm i \epsilon_{i j}\, k_2^j \over k_2^+}
\bigg\}
\ee 
It should be noted that our final state quark spinors are labeled by their momenta in a 
bar-ed frame in which their momenta are all longitudinal while momenta of the initial 
state quark spinors are in the original center of mass frame . Therefore we need to transform 
all the final state momenta and spinors and write them in the original frame, this can be 
formally done using the kinematical generators of Poincar\'e group in the light front 
form~\cite{cb} and light front spinors~\cite{dixon} (note that our normalization of 
the light front spinors differs from \cite{dixon} and is the same as \cite{ahjt}). Here 
we denote by $k_1 \, (p_1)$ the momentum vector constructed from the components 
of $\bar{k} \, (\bar{p}_1)$ in the original frame. We then have 

\bea
\mathcal{N}_{1 , b}^{+} &=& 
p^+ \, \sqrt{{q^+\over p^+}}\left\{2\, A^-_b (x) -  A^i_b (x) 
\bigg[{k_{1 i} - i \epsilon_{i j} \, k_1^j \over q^+ }
+ {  k_i + i \epsilon_{i j}\, k^j \over p^+}\bigg]\right\}
\nn
\mathcal{N}_{2 , b}^{+} &=&  {p^+ \over q_\perp^2} \sqrt{{q^+ \over p^+}}  
\bigg\{
\Big(1 + {q^+\over p^+}\Big) \, q_\perp\cdot A_\perp^b (x) + i \,
\Big(1 - {q^+\over p^+}\Big) \,  \epsilon^{i j} \, q_i\,  A^b_j (x) \bigg\} 
\nn
\mathcal{N}_{3 , b}^{+} &=& \mathcal{N}_2^{+ , b} (q_i \rightarrow p_{1 i}) \nn
&=& {p^+ \over p_{1 \perp}^2} \sqrt{{q+ \over p^+}} \bigg\{
\Big(1 + {q^+\over p^+}\Big) \, p_{1 \perp}\cdot A_\perp^b (x) + i \,
\Big(1 - {q^+\over p^+}\Big) \,  \epsilon^{i j} \, p_{1 i}\,  A^b_j (x) \bigg\} 
\eea
To facilitate comparison with the contribution of the eikonal term $\sim p^+ \sim \sqrt{s}$, we have extracted an explicit factor of $p^+$ above. We also note the appearance of imaginary term $i \epsilon^{i j}$ in the helicity amplitudes which will lead to spin asymmetry. Next we square the helicity amplitudes which will be needed for calculation of the scattering cross section. This is straightforward and gives 
\bea
|\mathcal{N}_1^{+ ,\, b \, c}|^2 &=&
4 p^+ q^+ \, A^-_b (x) \, A^-_c (y) \nn
&-& 
2 A^-_b (x) \, [q^+\, l_\perp + p^+ l_{1 \perp}]\cdot A^c_\perp (y) 
- 2 A^-_c (y) \, [q^+\, k_\perp + p^+ k_{1 \perp}]\cdot A^b_\perp (x) 
\nn
&+&
k_{1 \perp}\cdot A^b_\perp (x) \, l_\perp \cdot A^c_\perp (y) 
+ k_\perp\cdot A^b_\perp (x) \, l_{1 \perp} \cdot A^c_\perp (y)
\nn
&+&
l_\perp\cdot A^b_\perp (x) \, k_{1 \perp} \cdot A^c_\perp (y)
+ l_{1 \perp}\cdot A^b_\perp (x) \, k_\perp \cdot A^c_\perp (y)  
\nn
&-&
[k_\perp \cdot l_{1 \perp} + l_\perp\cdot k_{1 \perp}] \,  
A^b_\perp (x) \cdot  A^c_\perp (y) 
\nn
&+&
{q^+ \over p^+} [k_\perp\cdot A^b_\perp (x) \, l_\perp \cdot A^c_\perp (y) - 
l_\perp\cdot A^b_\perp (x) \, k_\perp \cdot A^c_\perp (y) + 
k_\perp \cdot l_\perp \, A^b_\perp (x) \cdot A^c_\perp (y)]\,
\nn
&+&
{p^+ \over q^+} [k_{1 \perp}\cdot A^b_\perp (x) \, l_{1 \perp} \cdot A^c_\perp (y) - 
l_{1 \perp}\cdot A^b_\perp (x) \, k_{1 \perp} \cdot A^c_\perp (y) + 
k_{1 \perp} \cdot l_{1 \perp} \, A^b_\perp (x) \cdot A^c_\perp (y)]
\nn
&+&
i\, \epsilon_{i j} \,\bigg[ 
2 [p^+ l_1^i - q^+ l^i]\, A^-_b (x) \, A_c^j (y)
- 2 [p^+ k_1^i - q^+ k^i]\, A^j_b (x) \, A^-_c (y) \nn
&+&
[l^i k_{1 \perp}  - l_1^i k_\perp 
  + {q^+ \over p^+} \, l^i k_\perp 
   - {p^+ \over q^+} \, l_1^i k_{1 \perp}] \cdot A_\perp^b (x)\, A^j_c (y)
\nn
&+&
[k_1^i l_\perp - k^i l_{1 \perp}  
  - {q^+ \over p^+} \, k^i l_\perp 
   + {p^+ \over q^+} \, k_1^i l_{1 \perp}]\,\cdot A_\perp^c (y) \, A^j_b (x)
\bigg]
\eea
and
\bea
|\mathcal{N}_2^{+ , \, b \, c}|^2 &=&  {q^+ \over p^+} \, {1 \over q_\perp^4} 
\bigg\{\bigg[(4 p^+\, q^+)\, q_\perp \cdot A_\perp^b (x) \,
 q_\perp \cdot A_\perp^c (y)  +
(p^+ - q^+)^2 \, q_\perp^2\, A_\perp^b (x) \cdot A_\perp^c (y)\bigg]
 \nn
&+&
i\, \epsilon^{i j} \, [(p^+)^2 - (q^+)^2]\, \bigg[
q_i \, A^b_j (x) \, q_\perp \cdot A_\perp^c (y)  -
q_i \, A^c_j (y) \, q_\perp \cdot A_\perp^b (x) \bigg] 
\bigg\} 
\eea
and
\bea
|\mathcal{N}_3^{+ , \, b \, c}|^2 &=& {q^+ \over p^+} 
{1 \over p_{1\perp}^2 \, p_{2\perp}^2}
\bigg\{\bigg[(p^+ + q^+)^2\, p_{1 \perp} \cdot A_\perp^b (x) \,
 p_{2 \perp} \cdot A_\perp^c (y) \nn
&-& \!\!(p^+ - q^+)^2 \left[p_{2 \perp}\!\cdot\! A_\perp^b (x) \,
p_{1 \perp}\!\cdot\! A_\perp^c (y) - p_{1 \perp}\!\cdot\! p_{2 \perp} 
A_\perp^b (x)\!\cdot\! A_\perp^c (y)\right]
\bigg]
 \nn
&+&
\!\! i\, \epsilon^{i j} \, [(p^+)^2 - (q^+)^2]\, \bigg[
p_{1 i} \, A^b_j (x) \, p_{2 \perp} \!\cdot\! A_\perp^c (y)  -
p_{2 i} \, A^c_j (y) \, p_{1 \perp} \!\cdot\! A_\perp^b (x) \bigg] 
\bigg\} 
\eea
There are also the off-diagonal terms in the amplitude squared which are not written out 
here to save space and specially since they add nothing qualitatively new to the diagonal terms included here. Using relations (\ref{eq:hel-plus-minus}) it should be clear that all the imaginary 
terms proportional to $i \epsilon^{i j}$ above will cancel between contribution 
of positive and negative helicity quarks to the unpolarized scattering cross section. 
On the other hand, if one is interested in spin asymmetry (in this case the so called double asymmetry) 
\be
A_{L L} \equiv { 
d \sigma^{+ +} - d \sigma^{- -} \over  d \sigma^{+ +} + d \sigma^{- -}  
}
\label{:eq:spinasymmetry}
\ee
the real terms in the helicity amplitude (squared) cancel and the $i \epsilon^{i j}$ terms will survive and give a non-zero spin asymmetry. The appearance of this non-zero spin asymmetry may seem confusing at first since we have said nothing about the helicity of the partons in the target which radiate the gluon field from which the quark is scattering. While the exact form of the asymmetry will depend on the assumptions made about the target partons, it is clear that the color field $A^\mu$ radiated by the target partons (large $x$ gluons in our case) will "know" about helicity of the target parton which is radiating it, as is illustrated in~\cite{kps} (see eqs. 19 - 20). Sub-eikonal corrections  at small $x$ are known to contribute to various spin observables~\cite{smallxhelicity}. In our case, however, this spin asymmetry is generated at intermediate-large $x$ corresponding to larger transverse momentum of the scattered quark. 

Furthermore, non-eikonal corrections to high energy scattering at small $x$ have been shown to generate angular asymmetries in particle production at small $x$~\cite{aaa}. Very similarly the large $x$ corrections to eikonal scattering considered here results in angular asymmetries, with the important difference that the asymmetries generated in our formalism will be at higher transverse momenta than the standard small $x$ results. We emphasize that, unlike eikonal scattering, the projectile quark here loses energy (rapidity) which could be used to investigate beam rapidity loss and the limiting fragmentation phenomenon~\cite{limitfrag} in high energy collisions. Furthermore, our approach allow one to calculate ultra-high energy neutrino-nucleon cross sections~\cite{hj} which receive significant contributions from the small $x$ kinematic region but are dominated by a large hard scale $\sim m_{W, Z}$.

There are several issues that need to be investigated further and clarified before one can apply our results to phenomenology. For instance, it is easy to show that all the new non-eikonal contributions vanish at small $x$ and one recovers the standard eikonal (tree level CGC) results. Taking the high transverse momentum limit, on the other hand, one is tempted to disregard all the soft multiple scatterings.  However this matching with pQCD must be done carefully~\cite{dmxy} in order to avoid difficulties with gauge invariance and needs to be better understood. A study of this limit in DIS structure functions calculated using our results here is in progress.

It will also be interesting to go beyond our tree level process and consider radiation. The simplest case is radiation of a photon from either the initial or final state quark~\cite{photon}. This would allow one to investigate photon-jet angular correlations from low to high $p_t$ (from low to high $x$) and study appearance/disappearance of the away side peak and its dependence on the exact kinematics of the process. One can also investigate cold matter energy loss, from incoherent Bethe-Heitler to coherent LPM regimes at the same time using our approach. Calculating photon radiation would also serve as a warm up for the calculation of gluon radiation~\cite{gluon} which would then allow one to investigate di-jet production and angular correlations in the full transverse momentum range, unlike the CGC calculations where one is limited to the low $p_t$ range. Radiation of a real gluon will also be part of the one-loop corrections to our tree level result, which augmented with virtual corrections, would lead to scattering cross sections which generalize and unify the small $x$ expressions of "$k_t$-factorized" CGC~\cite{jimwlk,bk} with collinearly-factorized pQCD~\cite{dglap} approach at large $x$.    

\section*{Acknowledgments}
We acknowledge support by the DOE Office of Nuclear Physics
through Grant No.\ DE-FG02-09ER41620 and by PSC-CUNY through 
grant No. 62185-0050. We would like to thank Centre de Physique Th\' eorique 
of \' Ecole Polytechnique for their kind hospitality during the completion of this 
work and the CNRS for support. We also thank T. Altinoluk, D. Boer, S. Brodsky, 
M. Hentschinski, W. Horowitz, C. Lorc\'e, P. Lowdon, C. Marquet, Y. Mehtar-Tani, 
A.H. Mueller, M. Sievert, W. Vogelsang and specially L. Dixon and Yu. Kovchegov 
for very helpful discussions.


\begin{thebibliography}{99}

\bibitem{cgc-reviews}
  E.~Iancu and R.~Venugopalan,
  In *Hwa, R.C. (ed.) et al.: Quark gluon plasma* 249-3363
  doi:10.1142/97898127955330005
  [hep-ph/0303204].
  F.~Gelis, E.~Iancu, J.~Jalilian-Marian and R.~Venugopalan,
  Ann.\ Rev.\ Nucl.\ Part.\ Sci.\  {\bf 60}, 463 (2010)
  doi:10.1146/annurev.nucl.010909.083629
  [arXiv:1002.0333 [hep-ph]]. 
  J.~Jalilian-Marian and Y.~V.~Kovchegov,
  Prog.\ Part.\ Nucl.\ Phys.\  {\bf 56}, 104 (2006)
  doi:10.1016/j.ppnp.2005.07.002
  [hep-ph/0505052].
  H.~Weigert,
  Prog.\ Part.\ Nucl.\ Phys.\  {\bf 55}, 461 (2005)
  doi:10.1016/j.ppnp.2005.01.029
  [hep-ph/0501087].

\bibitem{eic}
  E.~C.~Aschenauer {\it et al.},
  arXiv:1602.03922 [nucl-ex].

\bibitem{nlo}
  G.~A.~Chirilli, B.~W.~Xiao and F.~Yuan,
  Phys.\ Rev.\ D {\bf 86}, 054005 (2012)
  doi:10.1103/PhysRevD.86.054005
  [arXiv:1203.6139 [hep-ph]], 
  Phys.\ Rev.\ Lett.\  {\bf 108}, 122301 (2012)
  doi:10.1103/PhysRevLett.108.122301
  [arXiv:1112.1061 [hep-ph]];
  B.~W.~Xiao and F.~Yuan,
  arXiv:1407.6314 [hep-ph];
  I.~Balitsky and G.~A.~Chirilli,
  Phys.\ Rev.\ D {\bf 77}, 014019 (2008)
  doi:10.1103/PhysRevD.77.014019
  [arXiv:0710.4330 [hep-ph]], 
  Phys.\ Rev.\ D {\bf 83}, 031502 (2011)
  doi:10.1103/PhysRevD.83.031502
  [arXiv:1009.4729 [hep-ph]]; 
  Z.~B.~Kang, I.~Vitev and H.~Xing,
  Phys.\ Rev.\ Lett.\  {\bf 113}, 062002 (2014)
  doi:10.1103/PhysRevLett.113.062002
  [arXiv:1403.5221 [hep-ph]];
  E.~Iancu, A.~H.~Mueller and D.~N.~Triantafyllopoulos,
  JHEP {\bf 1612}, 041 (2016)
  doi:10.1007/JHEP12(2016)041
  [arXiv:1608.05293 [hep-ph]];
  A.~M.~Stasto and D.~Zaslavsky,
  Int.\ J.\ Mod.\ Phys.\ A {\bf 31}, no. 24, 1630039 (2016)
  doi:10.1142/S0217751X16300398
  [arXiv:1608.02285 [hep-ph]]; 
A.~Kovner, M.~Lublinsky and Y.~Mulian,
  JHEP {\bf 1408}, 114 (2014)
  doi:10.1007/JHEP08(2014)114
  [arXiv:1405.0418 [hep-ph]]; 
  K.~Roy and R.~Venugopalan,
  arXiv:1911.04530 [hep-ph].

\bibitem{dhj} 
  A.~Dumitru, A.~Hayashigaki and J.~Jalilian-Marian,
  Nucl.\ Phys.\ A {\bf 765}, 464 (2006)
  doi:10.1016/j.nuclphysa.2005.11.014
  [hep-ph/0506308].

\bibitem{gsv} 
  V.~Guzey, M.~Strikman and W.~Vogelsang,
  Phys.\ Lett.\ B {\bf 603}, 173 (2004)
  doi:10.1016/j.physletb.2004.10.033
  [hep-ph/0407201].

\bibitem{eik-reviews}
  A.~Kovner and U.~A.~Wiedemann,
  In Hwa, R.C. (ed.) et al.: Quark gluon plasma 192-248
  doi:10.1142/9789812795533$-$0004
  [hep-ph/0304151];
  J.~Casalderrey-Solana and C.~A.~Salgado,
  Acta Phys.\ Polon.\ B {\bf 38}, 3731 (2007)
  [arXiv:0712.3443 [hep-ph]].

\bibitem{jjm-largex}
  J.~Jalilian-Marian,
  Phys.\ Rev.\ D {\bf 96}, no. 7, 074020 (2017)
  doi:10.1103/PhysRevD.96.074020
  [arXiv:1708.07533 [hep-ph]], 
  Phys.\ Rev.\ D {\bf 99}, no. 1, 014043 (2019)
  doi:10.1103/PhysRevD.99.014043
  [arXiv:1809.04625 [hep-ph]].

\bibitem{martin} 
  M.~Hentschinski, A.~Kusina, K.~Kutak and M.~Serino,
  Eur.\ Phys.\ J.\ C {\bf 78}, no. 3, 174 (2018)
  doi:10.1140/epjc/s10052-018-5634-2
  [arXiv:1711.04587 [hep-ph]].
  M.~Hentschinski, A.~Kusina and K.~Kutak,
  Phys.\ Rev.\ D {\bf 94}, no. 11, 114013 (2016)
  doi:10.1103/PhysRevD.94.114013
  [arXiv:1607.01507 [hep-ph]].
  O.~Gituliar, M.~Hentschinski and K.~Kutak,
  JHEP {\bf 1601}, 181 (2016)
  doi:10.1007/JHEP01(2016)181
  [arXiv:1511.08439 [hep-ph]].
  I.~Balitsky and A.~Tarasov,
  JHEP {\bf 1606}, 164 (2016)
  doi:10.1007/JHEP06(2016)164
  [arXiv:1603.06548 [hep-ph]]; 
  JHEP {\bf 1510}, 017 (2015)
  doi:10.1007/JHEP10(2015)017
  [arXiv:1505.02151 [hep-ph]].

\bibitem{dixon}
  L.~J.~Dixon,
  hep-ph/9601359.

\bibitem{hybrid}
  A.~Dumitru and J.~Jalilian-Marian,
  Phys.\ Rev.\ Lett.\  {\bf 89}, 022301 (2002)
  doi:10.1103/PhysRevLett.89.022301
  [hep-ph/0204028], 
  Phys.\ Lett.\ B {\bf 547}, 15 (2002)
  doi:10.1016/S0370-2693(02)02709-0
  [hep-ph/0111357]; 
  J.~Jalilian-Marian and Y.~V.~Kovchegov,
  Phys.\ Rev.\ D {\bf 70}, 114017 (2004)
  Erratum: [Phys.\ Rev.\ D {\bf 71}, 079901 (2005)]
  doi:10.1103/PhysRevD.71.079901, 10.1103/PhysRevD.70.114017
  [hep-ph/0405266]; 
  C.~Marquet,
  Nucl.\ Phys.\ A {\bf 796}, 41 (2007)
  doi:10.1016/j.nuclphysa.2007.09.001
  [arXiv:0708.0231 [hep-ph]].

\bibitem{cb}
  K.~Y.~J.~Chiu and S.~J.~Brodsky,
  Phys.\ Rev.\ D {\bf 95}, no. 6, 065035 (2017)
  doi:10.1103/PhysRevD.95.065035
  [arXiv:1702.01127 [hep-th]].

\bibitem{ahjt}
  A.~Ayala, M.~Hentschinski, J.~Jalilian-Marian and M.~E.~Tejeda-Yeomans,
  Phys.\ Lett.\ B {\bf 761}, 229 (2016)
  doi:10.1016/j.physletb.2016.08.035
  [arXiv:1604.08526 [hep-ph]]; 
  Nucl.\ Phys.\ B {\bf 920}, 232 (2017)
  doi:10.1016/j.nuclphysb.2017.03.028
  [arXiv:1701.07143 [hep-ph]].

\bibitem{kps} 
  Y.~V.~Kovchegov, D.~Pitonyak and M.~D.~Sievert,
  JHEP {\bf 1710}, 198 (2017)
  doi:10.1007/JHEP10(2017)198
  [arXiv:1706.04236 [nucl-th]]. 

\bibitem{smallxhelicity}
  F.~Cougoulic and Y.~V.~Kovchegov,
  Phys.\ Rev.\ D {\bf 100}, no. 11, 114020 (2019)
  [arXiv:1910.04268 [hep-ph]]; 
  Y.~V.~Kovchegov and M.~D.~Sievert,
  Phys.\ Rev.\ D {\bf 99}, no. 5, 054032 (2019)
  doi:10.1103/PhysRevD.99.054032
  [arXiv:1808.09010 [hep-ph]]; 
  Y.~V.~Kovchegov, D.~Pitonyak and M.~D.~Sievert,
  Phys.\ Lett.\ B {\bf 772}, 136 (2017)
  doi:10.1016/j.physletb.2017.06.032
  [arXiv:1703.05809 [hep-ph]], 
  Phys.\ Rev.\ D {\bf 95}, no. 1, 014033 (2017)
  doi:10.1103/PhysRevD.95.014033
  [arXiv:1610.06197 [hep-ph]], 
  Phys.\ Rev.\ Lett.\  {\bf 118}, no. 5, 052001 (2017)
  doi:10.1103/PhysRevLett.118.052001
  [arXiv:1610.06188 [hep-ph]], 
  JHEP {\bf 1601}, 072 (2016)
  Erratum: [JHEP {\bf 1610}, 148 (2016)]
  doi:10.1007/JHEP01(2016)072, 10.1007/JHEP10(2016)148
  [arXiv:1511.06737 [hep-ph]].

\bibitem{aaa} 
  P.~Agostini, T.~Altinoluk and N.~Armesto,
  Eur.\ Phys.\ J.\ C {\bf 79}, no. 9, 790 (2019)
  doi:10.1140/epjc/s10052-019-7315-1
  [arXiv:1907.03668 [hep-ph]], 
  Eur.\ Phys.\ J.\ C {\bf 79}, no. 7, 600 (2019)
  doi:10.1140/epjc/s10052-019-7097-5
  [arXiv:1902.04483 [hep-ph]]; 
  T.~Altinoluk and A.~Dumitru,
  Phys.\ Rev.\ D {\bf 94}, no. 7, 074032 (2016)
  doi:10.1103/PhysRevD.94.074032
  [arXiv:1512.00279 [hep-ph]]; 
  T.~Altinoluk, N.~Armesto, G.~Beuf and A.~Moscoso,
  JHEP {\bf 1601}, 114 (2016)
  doi:10.1007/JHEP01(2016)114
  [arXiv:1505.01400 [hep-ph]]; 
  T.~Altinoluk, N.~Armesto, G.~Beuf, M.~Mart\'inez and C.~A.~Salgado,
  JHEP {\bf 1407}, 068 (2014)
  doi:10.1007/JHEP07(2014)068
  [arXiv:1404.2219 [hep-ph]].

\bibitem{limitfrag} 
  J.~Jalilian-Marian,
  Phys.\ Rev.\ C {\bf 70}, 027902 (2004)
  doi:10.1103/PhysRevC.70.027902
  [nucl-th/0212018]; 
  F.~Gelis, A.~M.~Stasto and R.~Venugopalan,
  Eur.\ Phys.\ J.\ C {\bf 48}, 489 (2006)
  doi:10.1140/epjc/s10052-006-0020-x
  [hep-ph/0605087].

\bibitem{hj} 
  E.~M.~Henley and J.~Jalilian-Marian,
  Phys.\ Rev.\ D {\bf 73}, 094004 (2006)
  doi:10.1103/PhysRevD.73.094004
  [hep-ph/0512220].

\bibitem{dmxy} 
  F.~Dominguez, C.~Marquet, B.~W.~Xiao and F.~Yuan,
  Phys.\ Rev.\ D {\bf 83}, 105005 (2011)
  doi:10.1103/PhysRevD.83.105005
  [arXiv:1101.0715 [hep-ph]].

\bibitem{photon} 
  F.~Gelis and J.~Jalilian-Marian,
  Phys.\ Rev.\ D {\bf 67}, 074019 (2003)
  doi:10.1103/PhysRevD.67.074019
  [hep-ph/0211363], 
  Phys.\ Rev.\ D {\bf 66}, 094014 (2002)
  doi:10.1103/PhysRevD.66.094014
  [hep-ph/0208141]; 
  J.~Jalilian-Marian,
  Nucl.\ Phys.\ A {\bf 753}, 307 (2005)
  doi:10.1016/j.nuclphysa.2005.02.156
  [hep-ph/0501222]. 

\bibitem{gluon}
  A.~Ayala, J.~Jalilian-Marian, L.~D.~McLerran and R.~Venugopalan,
  Phys.\ Rev.\ D {\bf 53}, 458 (1996)
  doi:10.1103/PhysRevD.53.458
  [hep-ph/9508302].

\bibitem{jimwlk}
J.~Jalilian-Marian, A.~Kovner, L.~D.~McLerran and H.~Weigert,
Phys.\ Rev.\  D {\bf 55}, 5414 (1997); 
J.~Jalilian-Marian, A.~Kovner, A.~Leonidov and H.~Weigert,
Nucl.\ Phys.\  B {\bf 504}, 415 (1997), 
Phys.\ Rev.\  D {\bf 59}, 014014 (1999), 
Phys.\ Rev.\  D {\bf 59}, 014015 (1999), 
Phys.\ Rev.\  D {\bf 59}, 034007 (1999),  
A.~Kovner, J.~G.~Milhano and H.~Weigert,
Phys.\ Rev.\  D {\bf 62}, 114005 (2000); 
A.~Kovner and J.~G.~Milhano,
Phys.\ Rev.\  D {\bf 61}, 014012 (2000); 
E.~Iancu, A.~Leonidov and L.~D.~McLerran,
Nucl.\ Phys.\  A {\bf 692}, 583 (2001),  
Phys.\ Lett.\  B {\bf 510}, 133 (2001);  
E.~Ferreiro, E.~Iancu, A.~Leonidov and L.~McLerran,
Nucl.\ Phys.\  A {\bf 703}, 489 (2002).

\bibitem{bk}
I. Balitsky, Nucl. Phys. {\bf B} {\bf 463}, 99 (1996); 
  Y.~V.~Kovchegov,
  Phys.\ Rev.\ D {\bf 60}, 034008 (1999)
  doi:10.1103/PhysRevD.60.034008
  [hep-ph/9901281].

\bibitem{dglap} 
G. Altarelli and G. Parisi, Nucl. Phys. {\bf B} {\bf
    126}, 298 (1977); V.N. Gribov, L.N. Lipatov, Sov. J. Nucl. Phys.
  {\bf 15}, 438 (1972); {\it ibid.} 675 (1972); Yu. Dokshitzer, Sov.
  Phys. JETP {\bf 46}, 641 (1977).




\end{thebibliography}
\end{document}